\documentclass[journal=jccte,manuscript=article]{achemso}

\usepackage[version=3]{mhchem}
\usepackage{color}
\usepackage{textcomp}
\usepackage{gensymb}
\usepackage{fixltx2e}
\usepackage{amsmath}
\usepackage{textgreek}
\usepackage{multirow}
\usepackage{booktabs}
\usepackage{rotating}
\usepackage{setspace}
\usepackage[symbol]{footmisc}

\author{Aalim S. Abdullah}
\affiliation{Kenneth S. Pitzer Theory Center and Department of Chemistry, University of California, Berkeley, CA 94720, USA}

\author{Yingze Wang}
\affiliation{Kenneth S. Pitzer Theory Center and Department of Chemistry, University of California, Berkeley, CA 94720, USA}

\author{Maximilian F.S.J. Menger}
\affiliation{Theoretische Chemie, Physikalisch-Chemisches Institut, University Heidelberg, INF 229, 69120 Heidelberg, Germany}

\author{Selim Sami}
\affiliation{Kenneth S. Pitzer Theory Center and Department of Chemistry, University of California, Berkeley, CA 94720, USA}
\email{s.sami@berkeley.edu}

\author{Teresa Head-Gordon}
\affiliation{Kenneth S. Pitzer Theory Center and Department of Chemistry, University of California, Berkeley, CA 94720, USA}
\alsoaffiliation{Chemical Sciences Division, Lawrence Berkeley National Laboratory, Berkeley, CA 94720, USA}
\alsoaffiliation{Departments of Bioengineering, Chemical and Biomolecular Engineering, University of California, Berkeley, CA 94720, USA}
\email{thg@berkeley.edu}

\title{Improved Treatment of 1-4 interactions in Force Fields for Molecular Dynamics Simulations}

\begin{document}

\vspace{-10mm}
\begin{abstract}
\vspace{-5mm}
\singlespacing
\noindent
Traditional force fields commonly use a combination of bonded torsional terms and empirically scaled non-bonded interactions to capture 1-4 energies and forces of atoms separated by three bonds in a molecule. While this approach can yield accurate torsional energy barriers, it often leads to inaccurate forces and erroneous geometries, and creates an interdependence between dihedral terms and non-bonded interactions, complicating parameterization and reducing transferability. In this paper, we demonstrate that 1-4 interactions can be accurately modeled using only bonded coupling terms, eliminating the need for arbitrarily scaled non-bonded interactions altogether. Furthermore by leveraging the automated parameterization capabilities of the Q-Force toolkit, we efficiently determine the necessary coupling terms without the need for manual adjustment. Our approach is first validated on a range of small molecule  systems, encompassing both flexible and rigid structures, and shows a significant improvement in force field accuracy, obtaining sub-kcal/mol mean absolute error for every molecule tested. We further extend the bonded-only model for 1-4 interactions to Amber ff14sb, CHARMM36, and OPLS-AA force fields to reproduce ab initio gas and implicit solvent  $\phi,\psi$ surfaces of alanine dipeptide.

\end{abstract} 

\newpage 
\singlespacing
\section{Introduction}

Accurately simulating molecular systems is fundamental to computational chemistry, enabling advancements in materials science, drug discovery, and biophysics. At the heart of classical simulations are force fields (FFs), physical models that estimate the potential energy of a (typically) non-reactive system based on both bonded terms: bond stretching, angle bending, and dihedral torsions, and non-bonded through space interactions including electrostatics, Pauli repulsion and dispersion, and other many-body effects such as polarization and charge transfer. Interactions between atoms separated by a single bond (1-2) or two bonds (1-3) are effectively captured by bond stretching and angle bending terms, respectively. Similarly, interactions between atoms separated by five or more bonds (1-5 and beyond) are typically represented solely by non-bonded terms. 

Bridging these two extremes are 1-4 interactions—interactions between atoms separated by three bonds which are critical for accurately modeling torsional energy barriers and molecular conformations. However, their treatment in traditional force fields is complex, relying on a combination of bonded dihedral terms and non-bonded interactions, creating challenges in parameterization, transferability, and accuracy. The hybrid treatment of 1-4 interactions introduces several significant limitations. The most critical issue arises from the spatial proximity of 1-4 atom pairs compared to typical non-bonded pairs. At these distances, charge penetration effects—where electron clouds overlap and influence electrostatic interactions—play a crucial role in modulating the interaction strength. However, standard non-bonded functions, such as the Lennard-Jones potential and Coulomb's law, do not account for charge penetration effects. As a result, the parameters governing these terms are either arbitrarily scaled by a constant or replaced with unique, smaller non-bonded parameters for 1-4 interactions compared to those used for other non-bonded interactions in the force field.

The more common approach, employed by force fields such as OPLS \cite{1988WLJorgensen, 2015MJRobertson}, AMBER \cite{1995WDCornell, 2004JWang}, CHARMM \cite{1998ADMacKerell, Vanommeslaeghe2010}, the Open Force Field class of FFs \cite{Qiu2021, Boothroyd2023}, and AMOEBA\cite{Ren2003, Ponder2010}, is to use a single scaling factor to modulate 1-4 non-bonded interactions. While this method is computationally simple and easy to implement, it is suboptimal for a diverse set of molecular interactions for several reasons. A single scaling parameter cannot capture the nuanced physical phenomena governing 1-4 interactions across different chemical environments, leading to reduced accuracy in force field predictions. While GROMOS \cite{1999WRPScott, 2011NSchmid} incorporates unique 1-4 parameters in specific cases, it defaults to scaled interactions for most systems. While this tailored approach has the potential to improve accuracy for certain molecules or functional groups, it does not fully address the underlying limitations of the standard Lennard-Jones and Coulomb potentials. These functional forms inherently lack the correct physics to represent interactions at short distances characteristic of 1-4 pairs. Although torsional barriers can be improved by fine-tuning 1-4 parameters, this does not guarantee accurate forces or a well-behaved potential energy surface (PES), which are critical for reliable simulations.

To address the inherent limitations of traditional methods for handling 1-4 interactions, we propose two possible pathways for improvement. The first involves treating 1-4 interactions entirely through non-bonded potentials with physically accurate functional forms. This would necessitate the development or adoption of advanced force fields where all non-bonded interactions, including those at short distances, are damped using physically motivated corrections, such as density overlap approaches \cite{kim1981dependence, wheatley1990overlap, van2016beyond}. Force fields like HIPPO \cite{Rackers2021} and CMM \cite{Heindel2024}, which incorporate density overlap to account for charge penetration and related effects, are promising candidates for this strategy. Such an approach could significantly reduce the reliance on torsional terms by inherently capturing the correct balance of interactions, including electrostatics, van der Waals forces, polarization, and charge transfer. However, while this approach holds promise, achieving the correct balance among these interaction components remains a considerable challenge. Subtle inaccuracies in modeling electrostatics, van der Waals interactions, or polarization effects could introduce errors that may still need to be compensated for by torsional terms. Additionally, the importance of polarization effects, particularly for 1-2 and 1-3 interactions, cannot be overlooked, as they can contribute significantly to the overall polarization energy at the 1-4 level \cite{Zgarbova2015,Tian2020}. Therefore, even with an advanced non-bonded potential, some torsional corrections may still be necessary to address residual discrepancies and ensure an accurate representation of the energy landscape.

The alternative approach involves treating the 1-4 interactions entirely as bonded terms. This bonded-only methodology not only circumvents the challenges associated with charge penetration effects discussed earlier but also decouples the non-bonded and torsional parameterization processes. This separation allows torsional terms to be directly optimized against quantum mechanical (QM) reference data without interference from non-bonded interactions. Such an approach necessitates handling the coupling between 1-4 interactions and 1-2 or 1-3 interactions—--effects that are traditionally, albeit imperfectly, addressed by non-bonded interactions in standard force fields. Fortunately, extensive research on coupled interaction terms, such as torsion-bond and torsion-angle coupling, provides a foundation for addressing these challenges. These terms were developed in the 1990s and implemented in force fields such as MM3 \cite{Allinger1989, Lii1989} and MMFF94 \cite{Halgren1996, Halgren1999}, CFF93 \cite{Hwang1994, Maple1994}. However, to our knowledge, all existing implementations still rely on non-bonded 1-4 interactions in addition to the coupling terms. Moreover, despite their potential, these coupling terms have so far not been widely adopted in common force fields. We believe the limited adoption of these coupling terms can be attributed to two primary challenges. First, there has been a lack of tools capable of systematically parameterizing this plethora of coupling terms. Second, generating the necessary QM reference data to calibrate these parameters accurately can also be a significant hurdle. Addressing these gaps requires an automated framework capable of deriving and optimizing these terms efficiently, as well as creating the necessary QM datasets.

In this work, we revisit the earlier concepts of coupling terms in order to propose a bonded-only treatment for 1-4 interactions, by implementing the coupling terms and developing the necessary protocols for generating and fitting QM data within Q-Force toolkit.\cite{Sami2021} Q-Force is a framework that enables a systematic, reproducible, and extensible approach to automated force field parameterization based on QM calculations which we harness for defining bonded coupling terms across 1-2, 1-3, and 1-4 coupling terms. We benchmark our results against traditional force fields, including OPLS and Sage, the polarizable force field AMOEBA, and the semi-empirical method xTB \cite{Bannwarth2019}. Our findings reveal substantial improvements in the accuracy of forces and energy surfaces, demonstrating that a bonded-only treatment of 1-4 interactions can effectively overcome the limitations of hybrid approaches. Additionally, by decoupling the parameterization of torsional and non-bonded terms, our method simplifies force field development, providing a clear and streamlined protocol that can be readily adopted by next-generation of both two-body and advanced force fields. To demonstrate the promise of this approach, we replace the 1-4 nonbonded interactions of AMBER ff14SB, CHARMM36, and OPLS-AA with our bonded-only model for alanine dipeptide in the gas phase and under implicit solvent, and find excellent agreement with the QM data.
\vspace{-4mm}

\section{Theory and Methods}
\vspace{-3mm}
In earlier work, we demonstrated that with the correct selection of bond, angle, and coupling terms, we can reproduce the PES in small molecules with very high accuracy\cite{Sami2024} which we briefly review here. In this work, we derive a fully bonded approach for torsion-containing molecules, with new torsion and torsion coupling terms introduced into Q-Force, which required a complete revamping of the QM data generation and the fitting procedures, which we also describe in this section.
\vspace{-3mm}

\subsection{Bond, angle, and bond-angle coupling terms}
\vspace{-2mm}
We continue to exploit the functional forms for bond, angle, and bond-angle couplings reported previously\cite{Sami2024} for the molecules in Figure 1, which we briefly recapitulate here.  

Bond distortions are modeled as a Morse bond stretching potential,
\begin{equation}
\label{eq:bond}
\textrm{V}_{\textrm{bond}}(b) = D_e \left(1 - e^{-\alpha (b - b_0)} \right)^2
\end{equation}
which captures the harmonic behavior near the equilibrium bond length $b_0$ as well as the anharmonic flattening at larger bond lengths. Here, $D_e$ represents the bond dissociation energy, determining the depth of the potential well, while $\alpha$ is the exponential decay constant that controls the width of the potential. The relation between $\alpha$ and the harmonic bond stretching force constant $K_b$ is given by:
\begin{equation}
\label{eq:bond2}
\alpha = \sqrt{\frac{K_b}{2D}}~
\end{equation}
In Q-Force, only the harmonic force constant $K_b$ is fitted,  whereas $D_e$ is looked up from a table containing experimental bond dissociation energies \cite{CRC_BDE_2016}. Coupling between bond lengths is captured by the bond-bond coupling term
\begin{equation}
\label{eq:bond-bond}
\textrm{V}_{\textrm{bond-bond}}(b, b') = K_{bb} (b - b_0)(b' - b'_0)
\end{equation}
where $K_{bb}$ is the coupling constant. This term accounts for energy changes arising from simultaneous distortions of two bonds that share a common atom. 

Angle bending is modeled using a cosine angle potential
\begin{equation}
\label{eq:angle}
\textrm{V}_{\textrm{angle}}(\theta) = K_\theta (\cos(\theta) - \cos(\theta_0))^2
\end{equation}
where $\theta_0$ is the equilibrium angle, and $K_\theta$ is the associated force constant.\cite{Dateo1994} This cosine-based formulation introduces accurate anharmonicity to the potential without resorting to higher-order polynomials as is done in some force fields such as AMOEBA.\cite{Ponder2010,Liu2019} For angular coupling, the angle-angle interaction term is given by
\begin{equation}
\label{eq:angle-angle}
\textrm{V}_{\textrm{angle-angle}}(\theta, \theta') = K_{\theta \theta} (\cos(\theta) - \cos(\theta_0)) (\cos(\theta') - \cos(\theta'_0))
\end{equation}
where $K_{\theta \theta}$ quantifies the interaction strength between two adjacent angles that share at least two common atoms.

Finally, the interaction between bond stretching and angle bending is expressed by
\begin{equation}
\label{eq:bond-angle}
\textrm{V}_{\textrm{bond-angle}}(b, \theta) = K_{b\theta} (b - b_0) (\cos(\theta) - \cos(\theta_0))
\end{equation}
where $K_{b\theta}$ is the corresponding force constant. This term is applied to all bond-angle pairs that share two common atoms.
\vspace{-3mm}

\subsection{Torsion terms and their coupling}
\vspace{-2mm}
For flexible torsions, which are torsions with multiple minima, we use an adapted version of the CFF93 model\cite{Hwang1994, Maple1994}. Dihedral torsions are described by the Fourier functional form
\begin{equation}
\label{eq:dihedral}
V_{torsion}(\phi) =  \sum_{n=1}^{4} K_{n, \phi} (1+\cos(n\phi)) 
\end{equation}
where $n$ denotes the periodicity, and $K_{n, \phi}$ represents the torsional force constant. To account for cross-interactions involving torsional angles, we include dihedral-bond, dihedral-angle and dihedral-angle-angle coupling terms. 

The dihedral-bond interaction is described by
\begin{equation}
\label{eq:dihedral-bond}
V_{torsion-bond}(b, \phi) =  \sum_{n=1}^{4} K_{n, b\phi} (b - b_0) (1+\cos(n\phi) 
\end{equation}
where $K_{n, b\phi}$ represents the coupling strength between bond distortions and torsional motions. This term applies to torsion-bond pairs sharing at least one atom. Similarly, the dihedral-angle interaction is modeled by
\begin{equation}
\label{eq:dihedral-angle}
V_{torsion-angle}(\theta, \phi) =  \sum_{n=1}^{4} K_{n, \theta\phi} (cos{\theta} - cos{\theta_0} (1+cos{n\phi})
\end{equation}
with $K_{n, \theta\phi}$ represents the coupling constant for torsion-angle pairs sharing at least two atoms. A higher-order coupling term for simultaneous angle deformations is introduced as:
\begin{equation}
\label{eq:dihedral-angle-angle}
\textrm{V}_{\textrm{torsion-angle-angle}}(\theta, \theta', \phi) = K_{\theta \phi \phi'} (\theta - \theta_0)(\theta' - \theta'_0) \cos \phi
\end{equation}
This term only applies to adjacent angles forming a torsion, such as the A-B-C and B-C-D angles of a torsion A-B-C-D, with $K_{\theta \phi \phi'}$ representing the coupling strength. This term has been shown to significantly contribute to both relative energies and forces, and our results confirm its importance in improving the accuracy of the force field.

For rigid torsions, defined by a single energy minimum (e.g., double bonds in ethene or aromatic systems like benzene), the same functional forms are employed, but only the $n=2$ term is retained in the torsion, torsion-bond, and torsion-angle terms. For planar centers with three neighbors, such as the central carbon in acetaldehyde or aromatic carbons in benzene, an improper dihedral term is added, represented by a harmonic potential:
\begin{equation}
\label{eq:improper}
\textrm{V}_{\textrm{improper}}(\phi) = K_{imp} \phi^2 
\end{equation}
This term ensures planarity by penalizing out-of-plane distortions. All three combination of improper dihedrals are added to each planar center.
\vspace{-3mm}

\subsection{Q-Force parameterization strategy for small molecules}
QM reference data was generated using the $\omega$-B97X-V \cite{Mardirossian2014} density functional, one of the most accurate hybrid-GGA functionals available \cite{Mardirossian2017,Goerigk2017}, in conjunction with the def2-TZVPD \cite{Weigend2005}, basis set. This combination was selected as a good compromise between accuracy and cost. Q-Chem software was used for all QM calculations \cite{Epifanovsky2021}. The 1-4 force fields were parameterized with the QM data using the latest version of Q-Force\cite{Sami2021}, which will be discussed in detail in a separate paper. However, below are the steps that are relevant for this work:

\begin{itemize}
    \item CREST \cite{Pracht2024} is used to generate all conformers of the small molecule. 
    \vspace{-2mm}
    \item For each molecule a total of 200 frames, divided equally between all conformers, are taken from an xTB GFN2 trajectory at 500K. Energies and forces are computed at the reference QM level for these snapshots.
    \vspace{-2mm}
    \item Relaxed torsional scans are performed with the reference QM method, and energies and forces are computed at the optimized structures.
    \vspace{-2mm}
    \item Force constants for all FF terms described above are fitted using Lasso Regression to minimize the difference between QM and MM energies and forces.
    \vspace{-1mm}
\end{itemize}

To validate our model and to assess its performance against other FF models, we generate the two following datasets:
\begin{itemize}
    \item For each molecule a total of 400 frames, divided equally between all conformers, are taken from an xTB GFN2 trajectory at 298K. 
    \vspace{-2mm}
    \item Relaxed torsional scans are performed at the FF level (i.e., structures correspond to FF constrained minimum).
\end{itemize}

To benchmark Q-Force against other FFs, we parameterized three additional FFs using well-established tools. The OPLS force field, a widely used standard FF, was parameterized using LigParGen \cite{Dodda2017}. Sage, a modern and QM-based FF developed within the Open Force Field Initiative, was parameterized using OpenForceField tools \cite{Wagner2024}. For a comparison with an advanced FF, we included AMOEBA, a polarizable FF that incorporates various bonded coupling terms\cite{Ponder2010}, parameterized using the Poltype2 toolkit \cite{Walker2022}. These three FFs have 1-4 non-bonded interactions that are scaled by different constants. The xTB GFN2 semi-empirical model\cite{Bannwarth2019} was also used as an additional benchmark to compare its performance against the FFs, as it uniquely occupies the space between FFs and QM methods in terms of both cost and accuracy. The OpenMM software platform was used for all FF evaluations.

We recognize that comparing Q-Force FF for small molecules, parameterized against the $\omega$-B97X-V reference, with other FFs optimized against different QM methods or objectives, does not constitute a perfectly fair comparison. For instance, Sage is parameterized against B3LYP-D3BJ with DZVP basis set. In order to estimate possible deviations arising from differences in QM references, we included the B3LYP-D3BJ \cite{Becke1992, Lee1988,Grimme2011} with TZVP basis set as a benchmark. Any error arising from differences in QM references should have similar magnitude to the deviation between these two QM methods. On the other hand, for AMOEBA, we adjusted Poltype settings so that any QM calculations it performs are done with $\omega$-B97X-V with the def2-TZVPD basis set.

\subsection{Protocol for 2D torsion scans for alanine dipeptide}
\vspace{-2mm}
For all Ramachandran plots, the minimum energy was subtracted from each point, and only energies within 7 kcal/mol of the minimum were contoured. All energies were graphed using cubic interpolation in python.
\begin{itemize}
    \item TorsionDrive\cite{Qiu2020} using xTB GFN2 was used to generate optimized structures at each $\phi$ and $\psi$ point. All angles were held fixed from $-165^{\circ}$ to $180^{\circ}$ in $15^{\circ}$ increments. (See Supplementary Figure S4) 
    \vspace{-2mm}
    \item The optimized coordinates from the xTB 2D scan were used as starting points for QM,AMBER ff14sb, CHARMM36, and OPLS-AA  optimizations while holding the corresponding $\phi$ and $\psi$ angles fixed.
    \vspace{-2mm}
    \item Converged QM points were used as starting points for Q-Force optimizations at the same fixed $\phi$ and $\psi$ angles.
    \vspace{-2mm}
    \item All 1-5 and beyond interactions from AMBER ff14SB, CHARMM36, and OPLS-AA were subtracted from the QM surface, and Q-Force used to fit the 1-2, 1-3, and 1-4 bonded terms described by Eqs (1)-(11). By 1-5 and beyond we refer to the Lennard Jones and Coulomb potentials used to treat atom pairs in the same molecule separated by 4 or more bonds.
    \vspace{-2mm}
    \item The OpenMM software platform\cite{Eastman2017} was used to perform energy minimizations for all force fields except for OPLS-AA for which GROMOS software package\cite{Christen2005} was used since OpenMM does not support OPLS-AA natively.
\end{itemize}

To generate the solvated Ramachandran surfaces we used Solvation Model Density (SMD) for our QM implicit solvent, and GBneck2 (GBN2) for our MM implicit solvent. SMD was chosen because it is a widely used continuum solvent model that leverages the entire solute electron density to accurately describe bulk electrostatic, and non-electrostatic contributions across a variety of solvents. GBN2 was chosen as it is the most recent version of the Generalized Born implicit solvent models that is available in OpenMM/AMBER, and it was designed with proteins and small organic molecules in mind. Their pairing ensures a state-of-the-art approach to implicit solvation models at the QM and MM level. QM energies were computed with the Q-Chem water setting for SMD solvent, while MM energies used GBN2 with $\epsilon = 80$.
\vspace{-3mm}

\section{Results}
\begin{figure}[H]
        \centering
\includegraphics[width=0.9\textwidth]{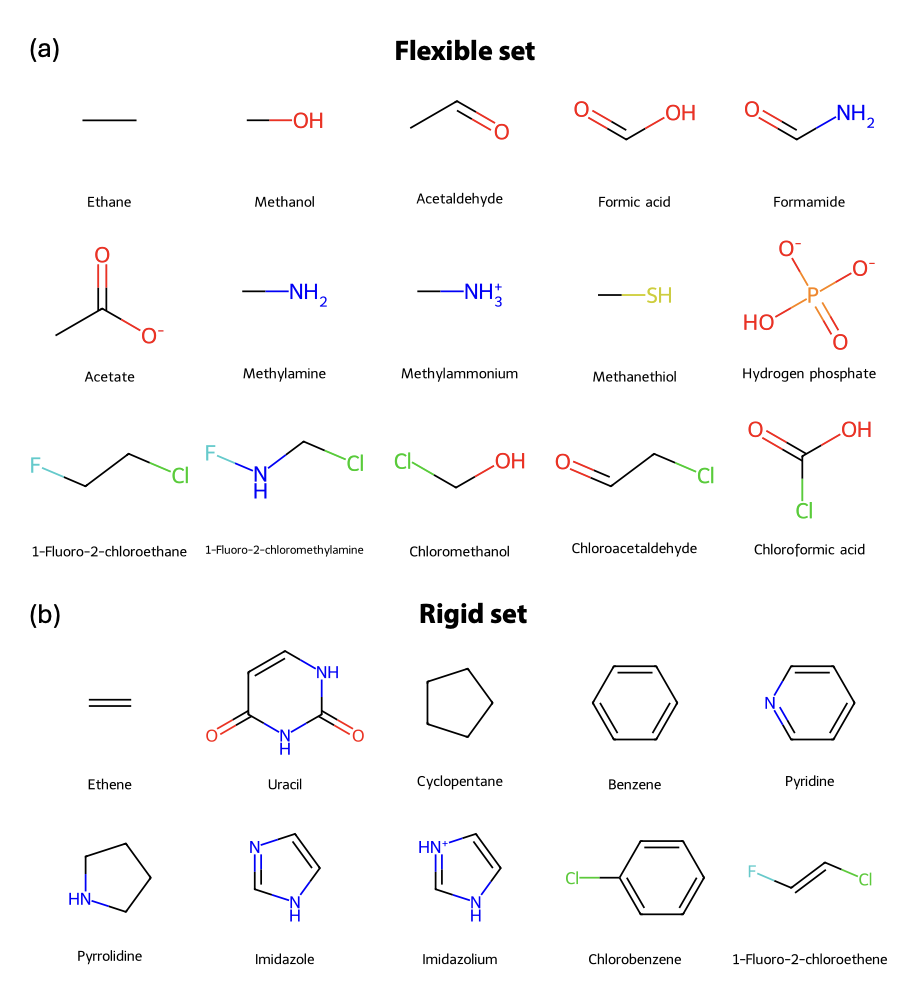}
        \caption{\textit{Flexible and rigid molecular datasets used in this study.} Molecules in the (a) flexible dataset contain a single fully rotatable torsion, whereas (b) the rigid set has molecules with multiple torsions that are restricted due to double bonds, conjugation or being in a ring.}
        \label{fig:molecules}
\end{figure}

\vspace{-2mm}
To test our approach for replacing 1-4 nonbonded interactions with torsions and their couplings, we evaluate its performance across a diverse set of small molecule systems encompassed in two datasets: the "flexible set" featuring molecules with flexible torsions (Figure \ref{fig:molecules}a) and the "rigid set" comprising of rigid molecules with either ring systems or double bonds (Figure \ref{fig:molecules}b). An important choice made here is that we only selected molecules with no discernible non-bonded interactions beyond 1-4, so that we can study the performance of the 1-4 bonded terms without having to choose a specific non-bonded FF, which could affect the interpretability of the results.

\subsection{Energy and force accuracy along the MD trajectory}
\begin{figure}[H]
        \centering
    \includegraphics[width=0.95\textwidth]{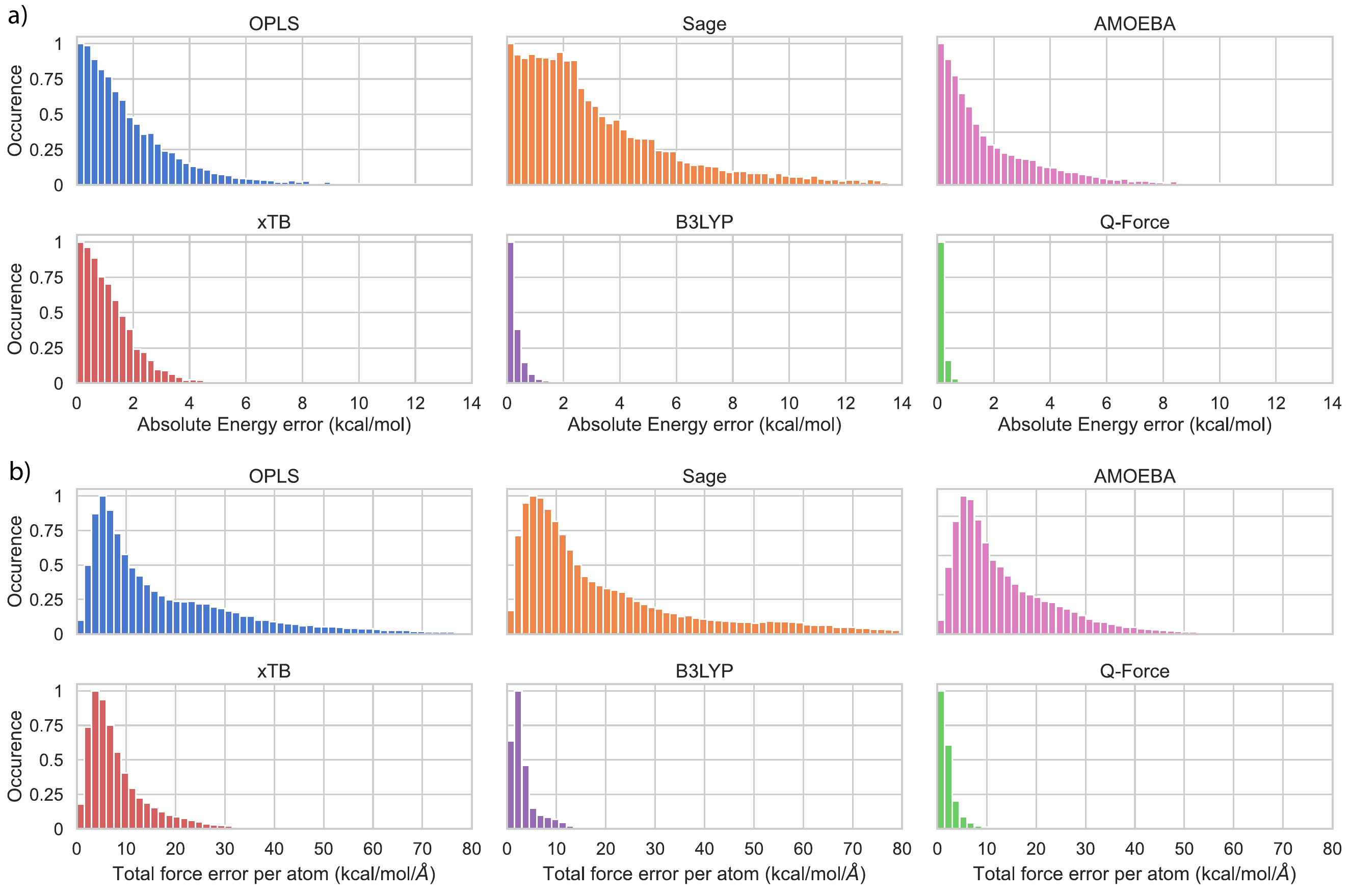}
        \caption{Absolute energy errors (a) and the norm of the vectorial force error per atom (b) with both datasets combined for all benchmarks in the study. The bin widths are 0.25 kcal/mol for energies and 1.5 kcal/mol/$\AA$ for forces. The occurrences are normalized to be 1 at maximum occurrence for each method.}
        \label{fig:energy_force_histograms}
\end{figure}

\vspace{-2mm}
As discussed in Methods, both the flexible and rigid datasets are generated by performing xTB GFN2 MD simulation at 298K and gathering 400 snapshots for each molecule and computing the energies and forces at the relevant geometries. The absolute energy error histograms are given in Figure \ref{fig:energy_force_histograms}A and the magnitude of the force error per atom is given in Figure \ref{fig:energy_force_histograms}B. Scatter plots for all the methods against the reference energies for both datasets are provided in Supplementary Figures S1-S3. Per molecule energy and force MAEs are also provided in Supplementary Table S1.

The mean absolute errors (MAEs) in energy and force for Q-Force are seen to be consistently an order of magnitude smaller on average compared to the other FFs across both datasets. Specifically, Q-Force achieves MAEs at or below -0.3 kcal/mol for energies and below -4 kcal/mol/$\mathrm{\mathring{A}}$ MAEs for forces across all molecules. In fact, Q-Force MAEs were lower than the deviation between $\omega$-B97X-V/def2-TZVPD and B3LYP-D3BJ/TZVP, showcasing its exceptional ability to reproduce the quantum mechanical (QM) reference data with high fidelity, likely to the accuracy limit of these QM methods. 

Comparing the two fixed charge FFs, Sage MAEs for energies are worse than OPLS by more than 1 kcal/mol, although Sage performs only a little worse than OPLS for forces across both datasets. The reason for the lower energy accuracy in Sage is due to the overestimation of the distortion energies, in which the average distortion energies corresponding to $\omega$-B97X-V, AMOEBA, OPLS and Sage are 6.9, 7.6, 9.4 and 11.7, respectively. For the flexible set, we also present the average values after removing the two worse outliers, hydrogen phosphate and chloroformic acid, in Supplementary Table 1
for which both fixed charge FFs perform especially poorly. In both cases Sage and OPLS significantly overestimate how repulsive these conformations are, which is particularly interesting because these are some of the test cases with the strongest Coulombic 1-4 interactions, indicating that the arbitrary scaling of 1-4 interactions might be even more problematic for such cases.

The polarizable AMOEBA FF does better than the fixed charge FFs across energies and forces for the flexible set, but the results are more mixed for the rigid set. OPLS outperforms AMOEBA in energies while AMOEBA does slightly better with forces in this case. AMOEBA's performance is particularly worse for ring systems, indicating that the automated parameterization strategy of Poltype2 toolkit \cite{Walker2022} could be improved in this regard. The semi-empirical xTB's performance for both energies and forces lies between the FFs and QM methods as expected, with $\sim$1 kcal/mol MAE in energies across both datasets. 
\vspace{-2mm}

\subsection{Torsional Profiles}
\vspace{-2mm}
Torsional scan profiles for the flexible dataset are shown in Figure \ref{fig:dihedral_scans} and the corresponding maximum energy and RMSD errors per molecule are given in Supplementary Table S2. These are relaxed scans performed at the constrained optimized geometry of each method. While it makes little difference for Q-Force or B3LYP-D3BJ/TZVP whether the energies are computed at the the reference relaxed or individual relaxed geometries, for other FF methods this results in larger deviations due to larger force errors as evidenced in the previous section. As such, we found it fairer to compare the energies at individually optimized geometries while investigating the correctness of the structures, hence the forces, by looking at the RMSD of geometries compared to the $\omega$-B97X-V/def2-TZVPD relaxed structure.

Q-Force consistently achieves the highest accuracy in reproducing torsional profiles, with maximum energy errors below 0.5 kcal/mol across all test cases relative to the QM reference. The RMSD in structures were also exceptionally low and in line with the QM reference; again Q-Force had smaller deviations to the reference than B3LYP-D3BJ/TZVP, indicating once again that it can reproduce the reference QM method exceptional fidelity. Among the Fixed charge force fields, Sage and OPLS both typically exceeded 1.5 kcal/mol errors in barrier heights, while performing worse for more complicated molecules, resulting in average maximum errors of 3.7 and 4.5 kcal/mol. AMOEBA demonstrated better performance, with an average maximum error below 2 kcal/mol. However, for several molecules, AMOEBA’s torsional barriers deviated by more than 4 kcal/mol, indicating limitations in its parameterization for certain cases. In terms of RMSD in structures, OPLS performed better than both Sage and AMOEBA, while the latter two having very similar values to each other. Semi-empirical xTB's performance fell between FFs and B3LYP-D3BJ/TZVP for both energies and RMSDs, with errors in torsional barrier heights of 1.1 kcal/mol on average. 
\vspace{-2mm}

\begin{figure}[H]
        \centering
        \includegraphics[width=\textwidth]{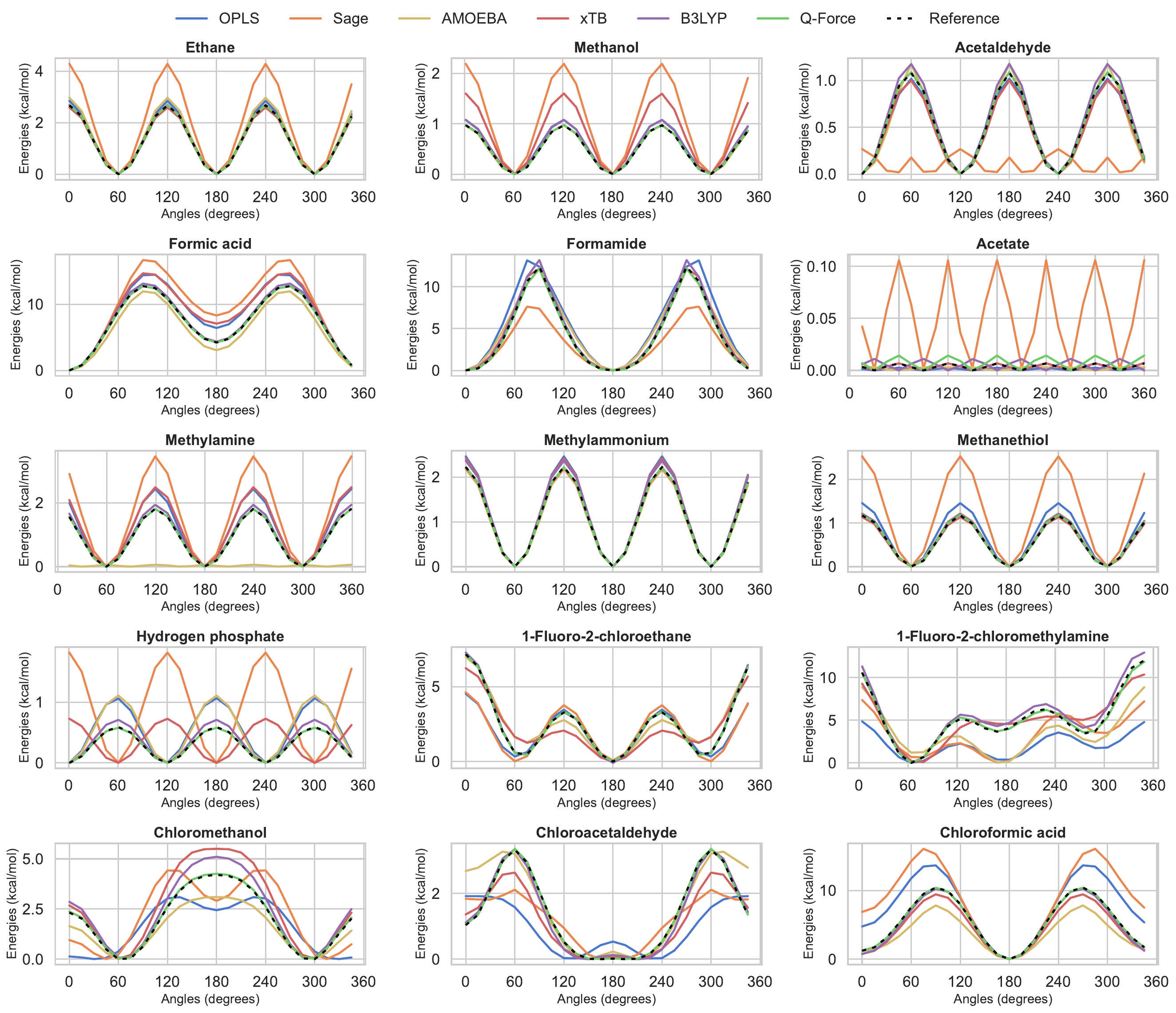}
        \caption{Torsional energy profiles corresponding to the molecules in the flexible dataset for each method. Geometries are relaxed for each method to their constrained minimum.}
        \label{fig:dihedral_scans}
\end{figure}

\subsection{Replacing 1-4 Non-bonded Interactions in Protein Force Fields}
\vspace{-2mm}
To evaluate the efficacy and validity of our new approach to treating 1-4 interactions for any type of force field, we consider the two dimensional \(\phi/\psi\) scans of alanine dipeptide. Alanine dipeptide is considered to be a building block of protein models, and the ability to correctly calculate its Ramachandran surface is key to elucidating if a force field is well suited to model conformational changes in proteins. Furthermore, unlike the molecules from the flexible and rigid datasets, alanine dipeptide contains numerous 1-5 and beyond (1-5+) interactions. 

We use Q-Force to fit the 1-2, 1-3, and 1-4 bonded terms and their couplings to the ab initio $\phi,\psi$ surface ($\omega$-B97X-V/def2-TZVPD) after subtracting the 1-5+ nonbonded interactions for each parent FF (OPLS, Amber, Charmm). It is important to note that we do not include the dihedral potentials from AMBER ff14sb, CHARMM36, and ff14sb, and the Coulomb and LJ parameters of the parent force field do not change. After fitting, the final
\begin{figure}[H]
  \centering
  \includegraphics[width=0.99\textwidth]{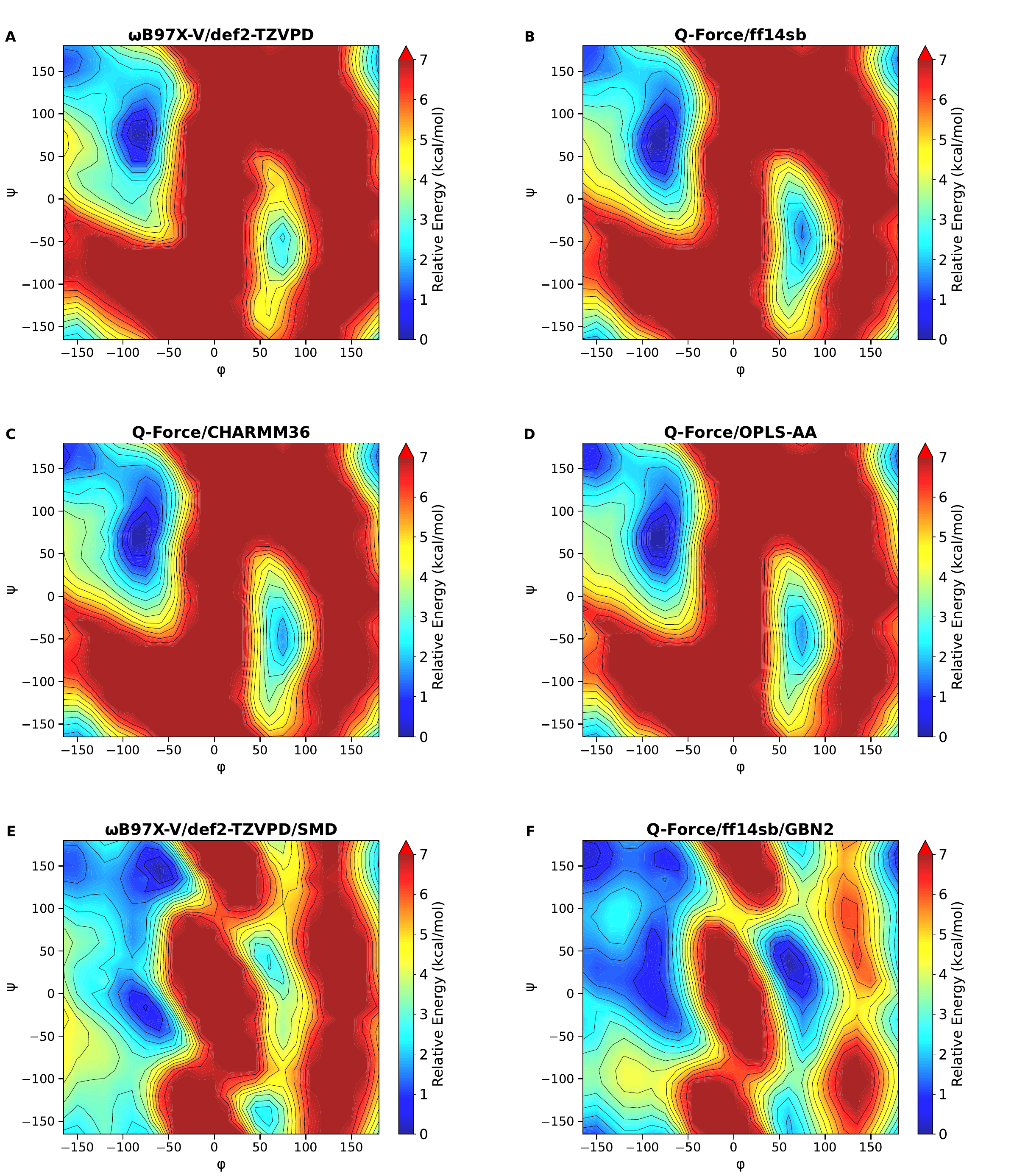}
  \caption{Ramachandran \(\phi/\psi\) potential energy surfaces for alanine dipeptide. 
  (A) $\omega$-B97X-V/def2-TZVPD, Q-Force for bonded and with (B) ff14sb 1-5+ nonbonded interactions, (C) CHARMM36 nonbonded 1-5+ interactions, (D) OPLS-AA nonbonded 1-5+ interactions. (E) $\omega$-B97X-V/def2-TZVPD with SMD and (F) Q-Force with ff14sb nonbonded interactions and GBN2 implicit solvent; energies are relative to each surface’s minimum to enable the comparison.}
  \label{fig:pes-side-by-side}
\end{figure}
\noindent
Q-Force/FF model is the sum of bond, angle, and torsion terms and their couplings, "embedded" into the unchanged non-bonded terms from each force field using their native parameters, but without any 1-4 nonbonded interactions.

Figure \ref{fig:pes-side-by-side}A shows the QM relaxed \(\phi/\psi\) surface of alanine dipeptide in the gas phase. This is compared to the Q-Force derived FFs with the bonded-only functional form, also in the gas phase, as seen in Figures \ref{fig:pes-side-by-side}B-D. The newly fitted maps reproduce the QM basin locations and barrier topography very well across the whole \((\phi,\psi)\) surface without need for 1-4 non-bonded interactions. Averaged across the 3 different force fields representing the 1-5+ interactions, the newly fitted 1-2, 1-3, and 1-4 terms yields very good accuracy, with a mean absolute error (MAE) \(\approx 0.3\) kcal/mole, root mean squared error (RMSE) \(\approx 0.5\) kcal/mole, and absolute maximum error of \(|\Delta E|\approx 2.25\) kcal/mol at the highest energy points on the map whose energy scale is $\sim$ 100 kcal/mol as seen in Supplementary Table S3. Supplementary Table S4 provides error metrics for the Ramachandran energy surfaces between Q-Force and the reference MM force field itself, and yields a MAE of \(\approx 0.5\) kcal/mole, RMSE of \(\approx 0.9\) kcal/mole, and \(|\Delta E|\approx 4.0\) kcal/mol on the same large energy scale.  These deviations arise from the imperfect 1-5+ non-bonded potential from the fixed charge FFs, and indicate that they are unable to capture all electronic effects from the QM reference such as polarization, charge transfer, and penetration due to missing many-body interactions. 

As previous theory has shown\cite{Shang1994}, the molecular origin of the $\alpha-$helical and left-handed helix minima in the dipeptide Ramanchandran plot is the stabilization that the reaction field from water solvent imparts to the aligned peptide dipoles of these configurations, which is a repulsive interaction in the gas phase. In addition, the Simmerling group used implicit solvent models with QM to parameterize the $\phi,\psi$ plots of amino acids to deal with this polarization issue.\cite{Tian2020} This is born out as seen when comparing the gas phase QM result in Figure \ref{fig:pes-side-by-side}A with the QM and implicit solvent SMD model in Figure \ref{fig:pes-side-by-side}E. It also allows us to use Q-Force to create a bonded model but now in the presence of implicit solvent as seen in Figure \ref{fig:pes-side-by-side}F. In the solvated case the agreement with the QM-SMD surface is less good than the gas phase, with MAE \(\approx 1.02\) kcal/mole, RMSE \(\approx 1.33\) kcal/mole, and max|ΔE| \(\approx 3.69\). The major contributor to the larger deviation is in the choice of solvent model employed, i.e., SMD at the QM level versus GBN2 at the MM level. 
\vspace{-2mm}

\section{Discussion and Conclusions}
\vspace{-2mm}
The reliance of fixed charge and polarizable FFs on a combination of bonded torsional terms and scaled non-bonded interactions to approximate 1-4 interactions introduces significant errors, arising from the arbitrary scaling of non-bonded terms and the dependence of torsional parameters on non-bonded interactions. These FF limitations result in inaccuracies in both potential energy surfaces and the accompanying forces, reducing their predictive power, as evidenced by the energies, forces, and torsion barriers of traditional FFs of various classes as shown in this work. 

In contrast, the bonded-only model developed in this work circumvents these issues entirely for modeling 1-4 interactions. By replacing the hybrid treatment of torsional and non-bonded interactions with explicitly parameterized bonded coupling terms, we achieve exceptional accuracy in reproducing QM reference data across a diverse set of molecules, encompassing both flexible and rigid systems. The success of this approach highlights the importance of explicitly accounting for coupled bonded interactions, as we recently showed for bonds and angles\cite{Sami2024} and now extended to torsion-bond and torsion-angle, torsion-angle-angle terms, which have been underutilized in FF development in our view. 

An additional and important feature of this work is the improved development of Q-Force\cite{Sami2021} for parameterizing these new bond coupling potentials. Notably, the errors in Q-Force predictions were even smaller than the discrepancies between the two QM methods used as benchmarks ($\omega$-B97X-V/def2-TZVPD and B3LYP-D3BJ/TZVP), suggesting that Q-Force closely approaches the inherent accuracy limits of the QM reference method. Importantly, the Q-Force toolkit is agnostic to the QM reference and the same methodology can be used to create FFs parameterized against coupled cluster, MP2 or any other QM method.\cite{Sami2021} Finally, the automated parameterization capabilities of Q-Force provide a reproducible and scalable framework for FF development in this regard as shown with ff14sb, CHARMM36, OPLS-AA, and is adaptable to advanced FFs by the same protocol as well. 

In summary, this work establishes that the bonded-only model is a viable and seemingly better alternative for modeling 1-4 interactions in molecular simulations, and underscores its potential to be used by next-generation FFs using Q-Force. By decoupling torsional and non-bonded parameterization, this methodology simplifies FF development, reduces interdependencies between terms, and will enhance transferability. A natural next step is to evaluate the performance of the bonded-only approach in the greater realm of atom-typed force fields, so that they can conveniently be used in larger scale applications such as materials design and drug discovery applications.

\begin{acknowledgement}
S.S. thanks the Dutch Research Council (NWO) Rubicon grant (019.212EN.004) for fellowship support. A.S.A., Y.W. and T.H-G. acknowledges support from the U.S. National Science Foundation through Grant No. CHE-2313791. Computational resources were provided by Savio at UC Berkeley.
\end{acknowledgement}

\begin{suppinfo}
Supporting information: Error metrics, additional Ramachandran plots, and force field parameters.
\end{suppinfo}

\bibliography{citations}

\begin{tocentry}
\centering
\includegraphics{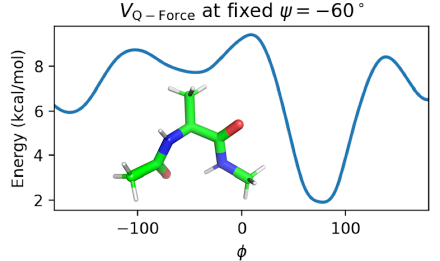}
\end{tocentry}

\end{document}


\section{Data Availability}
\noindent
All Q-Force force field parameters of the new functional form for 1-2, 1-3, and 1-4 interactions are available in a public github for all models examined in this work.\\

\noindent
https://github.com/THGLab/Q-Force\_alanine\_dipeptide\_forcefields

\newpage

\begin{figure}[!htb]
        \centering
        \includegraphics[width=0.8\textwidth]{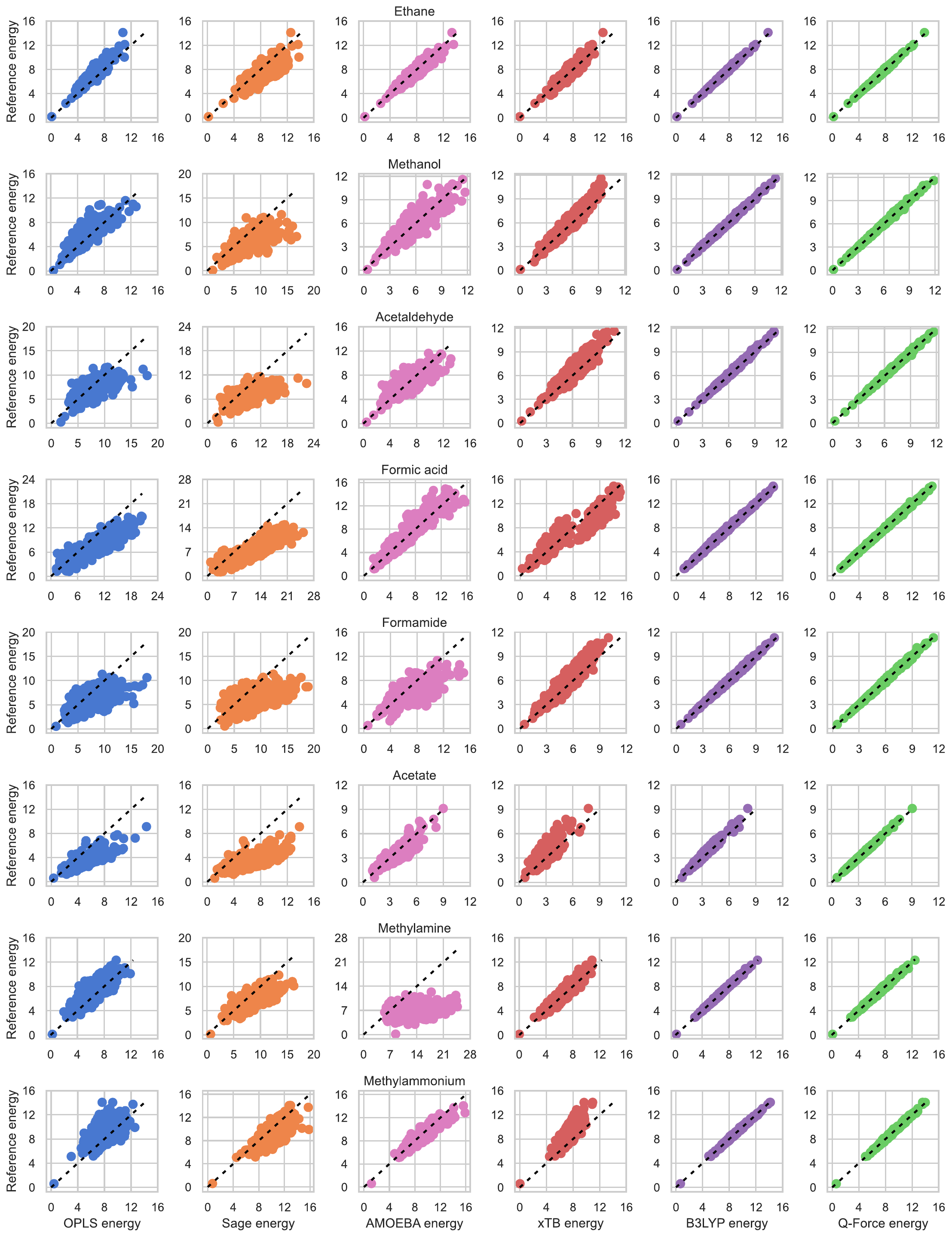}
        \caption{First half of the molecules in the flexible dataset. Per molecule scatter plot of the Reference (y axis) and benchmark (x axis) for all the methods in the study.}
        \label{fig:per_molecule_flex1}
\end{figure}

\newpage

\begin{figure}[!htb]
        \centering
        \includegraphics[width=0.8\textwidth]{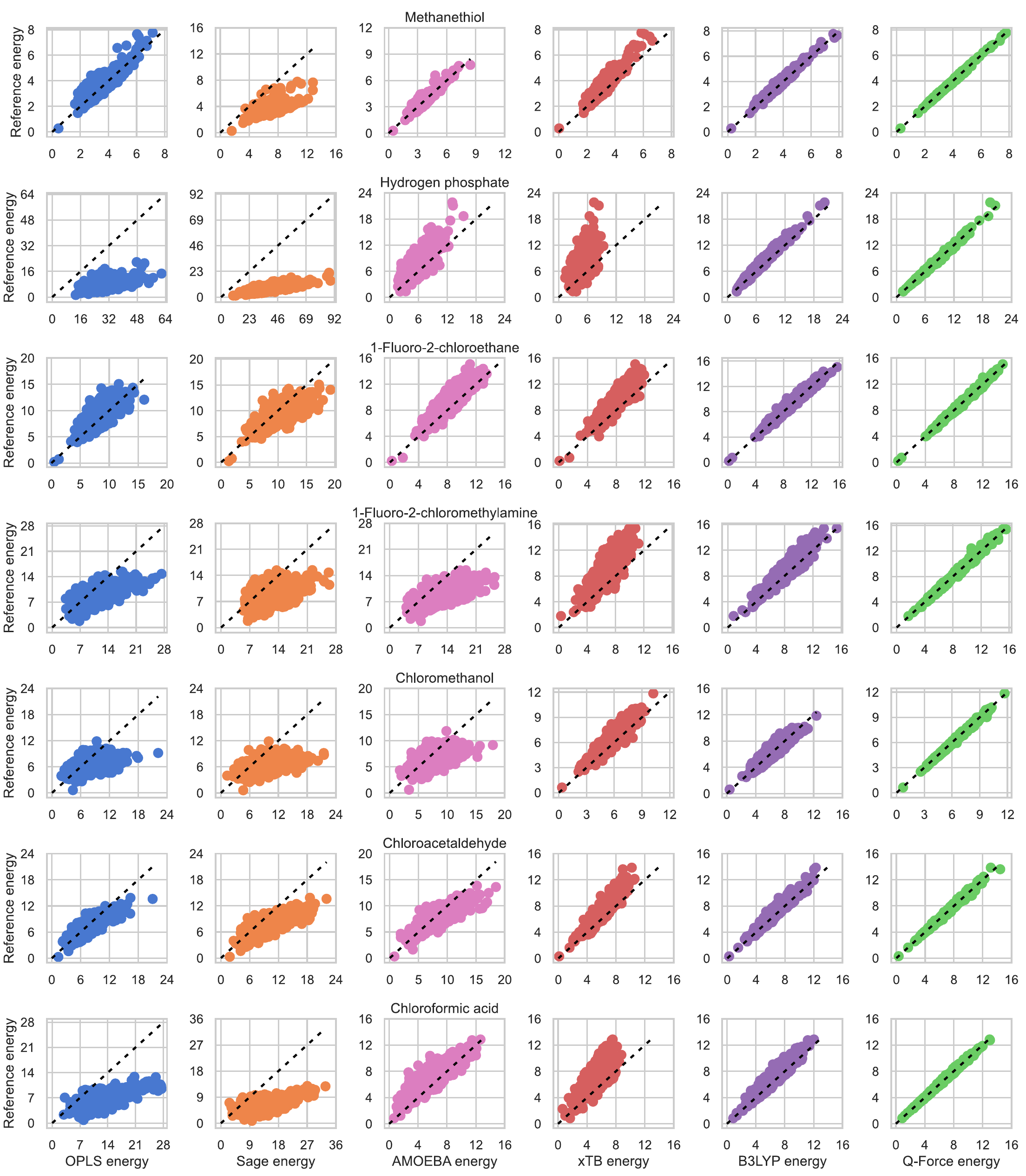}
        \caption{Second half of the molecules in the flexible dataset. Per molecule scatter plot of the Reference (y axis) and benchmark (x axis) for all the methods in the study.}
        \label{fig:per_molecule_flex2}
\end{figure}

\newpage

\begin{figure}[!htb]
        \centering
        \includegraphics[width=0.8\textwidth]{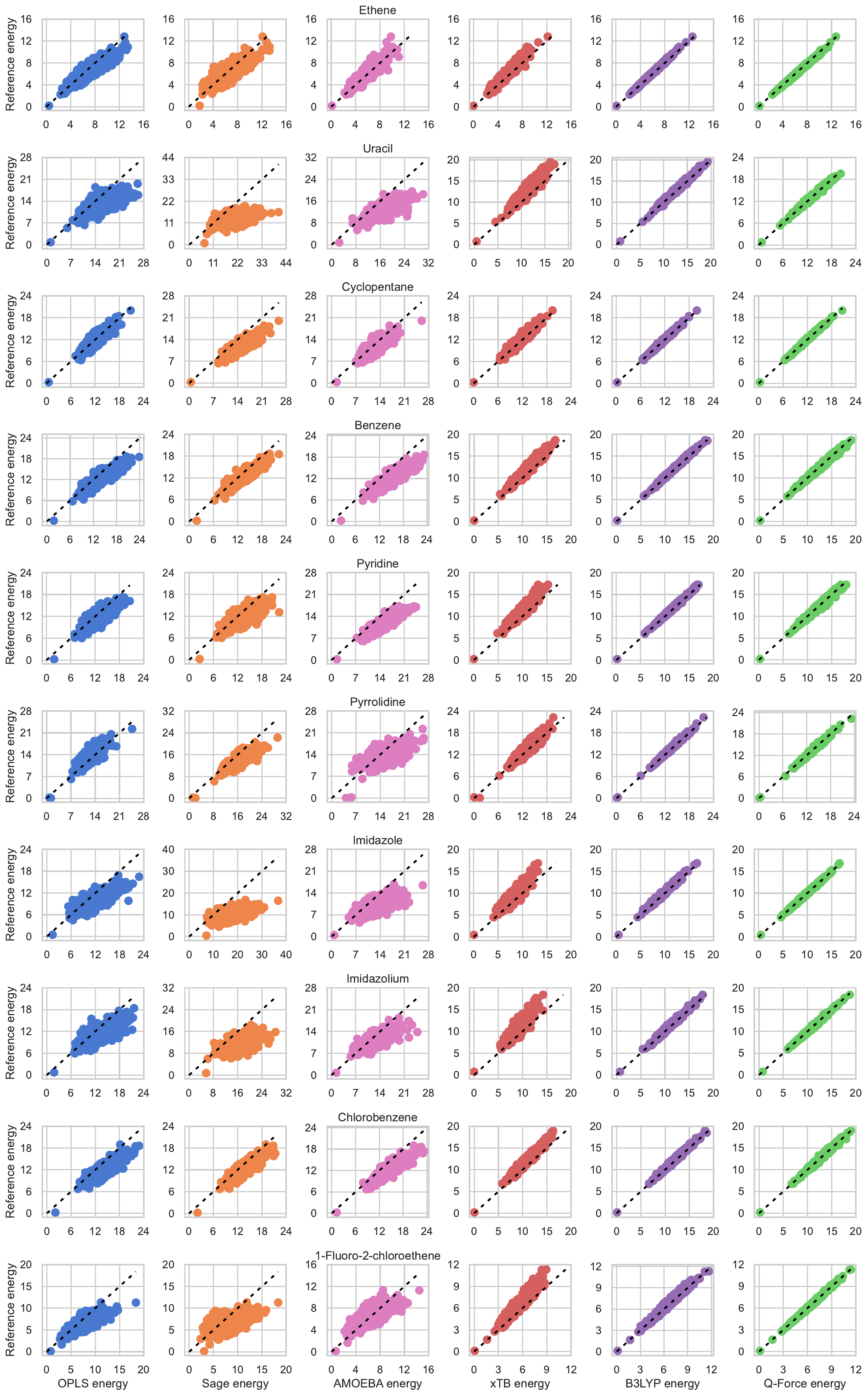}
        \caption{Molecules in the rigid dataset. Per molecule scatter plot of the Reference (y axis) and benchmark (x axis) for all the methods in the study.}
        \label{fig:per_molecule_rigid}
\end{figure}

\begin{figure}[!htb]
        \centering
        \includegraphics[width=0.99\textwidth]{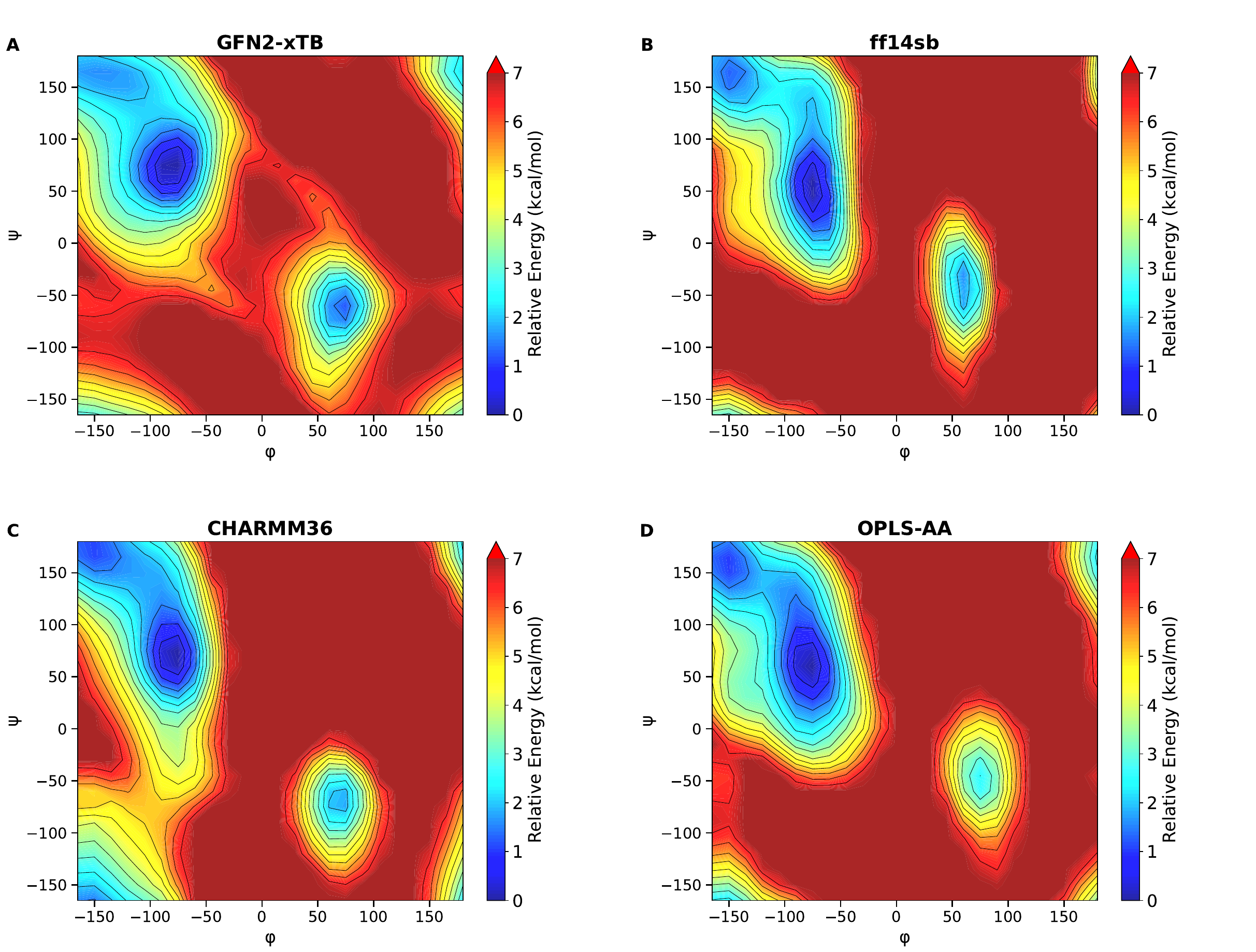}
        \caption{Ramachandran \(\phi/\psi\) potential energy surfaces for alanine dipeptide in the gas phase. (A) xTB GFN2, (B) ff14sb, (C) CHARMM36, and (D) OPLS-AA.}
        \label{fig:compare}
\end{figure}

\begin{sidewaystable}
\caption{MAE of energies and the mean of the norm of the vectorial force error per atom for every molecule in both datasets. Flexible/Average(-2) indicates that it is averaged across all molecules in the flexible set except the two worst performers.}
\label{table:energy_forces}
\begin{tabular}{lrrrrrrrrrrrr}
\multirow{2}{*}{} & \multicolumn{6}{c}{Energy MAEs (kcal/mol)} & \multicolumn{6}{c}{Force MAEs (kcal/mol/$\mathrm{\mathring{A}}$)} \\
\cmidrule(r){2-7} \cmidrule(l){8-13}
Molecule  & OPLS & Sage & Amoeba & xTB & B3LYP & Q-Force & OPLS & Sage & Amoeba & xTB & B3LYP & Q-Force \\
\midrule
Ethane & 0.988 & 1.378 & 0.521 & 0.441 & 0.123 & 0.084 & 6.66 & 8.30 & 5.53 & 5.19 & 1.78 & 1.30 \\
Methanol & 0.919 & 1.985 & 0.759 & 0.494 & 0.110 & 0.086 & 13.84 & 16.59 & 9.59 & 6.74 & 1.79 & 1.54 \\
Acetaldehyde & 1.482 & 2.196 & 0.824 & 0.693 & 0.154 & 0.083 & 20.02 & 24.47 & 13.15 & 7.77 & 1.84 & 1.42 \\
Formic acid & 3.069 & 4.779 & 0.934 & 1.351 & 0.139 & 0.100 & 36.02 & 35.84 & 14.85 & 13.89 & 2.07 & 2.23 \\
Formamide & 2.270 & 2.957 & 1.597 & 0.757 & 0.122 & 0.081 & 23.41 & 26.33 & 13.73 & 10.69 & 1.61 & 1.57 \\
Acetate & 1.786 & 3.553 & 0.602 & 0.752 & 0.269 & 0.048 & 17.95 & 21.47 & 7.79 & 14.02 & 3.33 & 1.25 \\
Methylamine & 0.933 & 1.440 & 6.796 & 0.476 & 0.114 & 0.117 & 10.24 & 11.19 & 19.05 & 5.82 & 1.57 & 1.80 \\
Methylammonium & 1.636 & 1.348 & 0.643 & 1.626 & 0.097 & 0.116 & 15.56 & 12.90 & 7.87 & 8.96 & 1.67 & 1.75 \\
Methanethiol & 0.449 & 2.962 & 0.285 & 0.492 & 0.147 & 0.048 & 5.20 & 12.35 & 4.80 & 6.60 & 3.78 & 0.89 \\
Hydrogen phosphate & 22.167 & 32.534 & 2.047 & 3.101 & 0.711 & 0.192 & 72.34 & 90.25 & 17.76 & 12.68 & 4.51 & 2.69 \\
1-F-2-Cl-ethane & 1.391 & 1.784 & 1.182 & 1.501 & 0.344 & 0.145 & 13.59 & 17.32 & 8.41 & 7.72 & 3.74 & 2.00 \\
1-F-2-Cl-MMA & 3.358 & 4.520 & 4.105 & 2.192 & 0.842 & 0.193 & 31.12 & 25.02 & 23.67 & 10.92 & 7.38 & 2.73 \\
Chloromethanol & 2.065 & 3.252 & 1.757 & 0.760 & 0.524 & 0.116 & 21.91 & 21.26 & 15.85 & 8.37 & 4.60 & 1.76 \\
Chloroacetaldehyde & 1.360 & 2.673 & 1.163 & 1.164 & 0.448 & 0.154 & 19.20 & 26.02 & 15.52 & 7.60 & 3.92 & 2.00 \\
Chloroformic acid & 6.947 & 8.312 & 1.431 & 1.809 & 0.626 & 0.118 & 46.52 & 48.34 & 20.60 & 15.59 & 4.66 & 2.21 \\
\midrule
Flexible/Average(-2) & 1.670 & 2.679 & 1.057 & 0.947 & 0.247 & 0.100 & 18.06 & 19.93 & 11.84 & 8.69 & 2.79 & 1.67 \\
Flexible/Average & 3.388 & 5.045 & 1.643 & 1.174 & 0.318 & 0.112 & 23.57 & 26.51 & 13.21 & 9.50 & 3.22 & 1.81 \\
\midrule
Ethene & 0.883 & 1.183 & 0.635 & 0.712 & 0.198 & 0.163 & 14.52 & 18.44 & 9.20 & 8.01 & 2.28 & 1.86 \\
Uracil & 3.001 & 7.537 & 4.666 & 1.731 & 0.302 & 0.237 & 20.82 & 34.35 & 17.99 & 9.32 & 3.52 & 2.95 \\
Cyclopentane & 0.879 & 3.108 & 1.524 & 0.637 & 0.191 & 0.137 & 6.80 & 9.39 & 10.09 & 4.90 & 2.61 & 1.43 \\
Benzene & 2.017 & 1.411 & 3.167 & 1.208 & 0.250 & 0.256 & 16.55 & 11.73 & 13.33 & 6.08 & 2.80 & 3.18 \\
Pyridine & 1.647 & 1.801 & 3.330 & 1.414 & 0.246 & 0.297 & 18.01 & 16.51 & 14.57 & 7.67 & 2.92 & 3.84 \\
Pyrrolidine & 1.330 & 2.918 & 3.127 & 0.807 & 0.244 & 0.257 & 10.19 & 10.79 & 14.56 & 5.63 & 2.59 & 2.07 \\
Imidazole & 2.299 & 7.095 & 2.533 & 1.266 & 0.338 & 0.117 & 21.77 & 34.16 & 17.60 & 9.90 & 4.50 & 1.67 \\
Imidazolium & 2.200 & 4.984 & 2.073 & 1.741 & 0.308 & 0.151 & 18.92 & 26.89 & 19.75 & 11.00 & 4.35 & 1.67 \\
Chlorobenzene & 2.346 & 1.588 & 2.407 & 1.324 & 0.295 & 0.225 & 17.21 & 12.53 & 12.71 & 5.98 & 3.30 & 3.29 \\
1-F-2-Cl-ethene & 1.438 & 2.261 & 1.062 & 0.927 & 0.248 & 0.070 & 15.98 & 27.42 & 15.48 & 7.48 & 3.34 & 1.42 \\
\midrule
Rigid/Average & 1.804 & 3.388 & 2.452 & 1.177 & 0.262 & 0.191 & 16.08 & 20.22 & 14.53 & 7.60 & 3.22 & 2.34 \\
\bottomrule
\end{tabular}
\end{sidewaystable}

\begin{sidewaystable}
\caption{Maximum energy (kcal/mol) and RMSD ($\mathrm{\mathring{A}}$) in structures for each molecule for torsional scans corresponding to Figure 3 in the main document. Average(-2) indicates that it is averaged across all molecules except the two worst performers.}
\label{table:scans}
\begin{tabular}{lrrrrrr|rrrrrr}
\multirow{2}{*}{} & \multicolumn{6}{c}{Max. energy errors (kcal/mol)} & \multicolumn{6}{c}{Max. RMSD ($\mathrm{\mathring{A}}$)} \\
\cmidrule(r){2-7} \cmidrule(l){8-13}
Molecule  & OPLS & Sage & Amoeba & xTB & B3LYP & Q-Force & OPLS & Sage & Amoeba & xTB & B3LYP & Q-Force \\
\midrule
Ethane & 0.23 & 1.99 & 0.32 & 0.18 & 0.03 & 0.01 & 0.015 & 0.028 & 0.010 & 0.015 & 0.003 & 0.004 \\
Methanol & 0.32 & 2.01 & 0.41 & 0.70 & 0.11 & 0.01 & 0.022 & 0.032 & 0.027 & 0.008 & 0.004 & 0.001 \\
Acetaldehyde & 1.29 & 1.81 & 0.42 & 0.19 & 0.11 & 0.01 & 0.031 & 0.067 & 0.061 & 0.018 & 0.005 & 0.002 \\
Formic acid & 3.87 & 6.53 & 1.29 & 3.19 & 0.39 & 0.15 & 0.042 & 0.084 & 0.061 & 0.026 & 0.006 & 0.006 \\
Formamide & 7.69 & 5.02 & 4.29 & 1.26 & 0.80 & 0.27 & 0.082 & 0.112 & 0.132 & 0.027 & 0.006 & 0.008 \\
Acetate & 0.71 & 1.99 & 0.25 & 0.48 & 0.05 & 0.01 & 0.026 & 0.060 & 0.024 & 0.017 & 0.005 & 0.002 \\
Methylamine & 0.77 & 2.09 & 8.21 & 0.81 & 0.13 & 0.03 & 0.033 & 0.034 & 0.290 & 0.017 & 0.003 & 0.004 \\
Methylammonium & 1.64 & 1.24 & 0.25 & 0.63 & 0.16 & 0.01 & 0.025 & 0.032 & 0.019 & 0.016 & 0.003 & 0.003 \\
Methanethiol & 0.32 & 1.90 & 0.23 & 0.17 & 0.09 & 0.01 & 0.028 & 0.043 & 0.026 & 0.018 & 0.006 & 0.002 \\
Hydrogen phosphate & 16.20 & 18.18 & 1.37 & 1.88 & 0.22 & 0.01 & 0.120 & 0.078 & 0.047 & 0.079 & 0.010 & 0.005 \\
1-F-2-Cl-ethane & 2.63 & 2.53 & 1.27 & 1.34 & 0.21 & 0.05 & 0.032 & 0.035 & 0.028 & 0.020 & 0.008 & 0.010 \\
1-F-2-Cl-MMA & 7.38 & 5.30 & 3.89 & 2.26 & 1.19 & 0.45 & 0.089 & 0.118 & 0.070 & 0.061 & 0.016 & 0.018 \\
Chloromethanol & 2.19 & 4.10 & 1.68 & 1.50 & 0.98 & 0.06 & 0.037 & 0.051 & 0.057 & 0.016 & 0.011 & 0.002 \\
Chloroacetaldehyde & 2.61 & 3.23 & 2.67 & 0.86 & 0.27 & 0.04 & 0.037 & 0.059 & 0.049 & 0.040 & 0.010 & 0.005 \\
Chloroformic acid & 7.63 & 10.18 & 2.92 & 1.00 & 0.47 & 0.10 & 0.085 & 0.137 & 0.100 & 0.032 & 0.011 & 0.009 \\
\midrule
Average(-2) & 2.43 & 3.06 & 1.31 & 0.84 & 0.23 & 0.04 & 0.038 & 0.055 & 0.045 & 0.021 & 0.006 & 0.004 \\
Average & 3.70 & 4.54 & 1.97 & 1.10 & 0.35 & 0.08 & 0.047 & 0.065 & 0.067 & 0.027 & 0.007 & 0.005 \\
\bottomrule
\end{tabular}
\end{sidewaystable}

\begin{table}
\centering
\small
\caption{Error metrics (kcal/mol) for Ramachandran Energy Surfaces of Q-Force Force Fields as compared to Reference QM.}
\label{table:ramachandran_errors}
\begin{tabular}{lrrr}
\toprule
Method & MAE & RMSE & $\max|\Delta E|$\\
\midrule
Q-Force/ff14sb & 0.310 & 0.532 & 2.457 \\
Q-Force/CHARMM36 & 0.296 & 0.484 & 2.055 \\
Q-Force/OPLS-AA & 0.318& 0.517& 2.108 \\
Q-Force/ff14sb/GBN2 & 1.016 & 1.331& 3.69 \\
\end{tabular}
\end{table}

\begin{table}
\centering
\small
\caption{Error metrics (kcal/mol) for Ramachandran Energy Surfaces of Q-Force Force Fields as compared to Reference MM Force Field.}
\label{table:ramachandran_errors}
\begin{tabular}{lrrr}
\toprule
Method & MAE & RMSE & $\max|\Delta E|$\\
\midrule
Q-Force/ff14sb & 0.598 & 1.046 & 4.312 \\
Q-Force/CHARMM36 & 0.681 & 1.107 & 4.151 \\
Q-Force/OPLS-AA & 0.452 & 0.783 & 3.510 \\
Q-Force/ff14sb/GBN2 & 1.219 & 1.674 & 4.960  \\
\end{tabular}
\end{table}